\documentclass[twocolumn,showpacs,preprintnumbers,amsmath,amssymb]{revtex4}
\usepackage{graphicx}
\usepackage{dcolumn}
\usepackage{bm}
\usepackage[latin1]{inputenc}

\begin{document}

\preprint{APS/123-QED}

\title{Interface Instability in Shear-Banding Flow\\}

\author{S. Lerouge}
\altaffiliation[Corresponding author ]{}
\email{slerouge@ccr.jussieu.fr}
\affiliation{%
Laboratoire Mati\`eres et Syst\`emes Complexes, Universit\'e Paris 7\\ 2 place Jussieu, 75251 Paris C\'edex 05}%
\author{M. Argentina}%
\affiliation{Institut Non Linéaire, Universit\'e de Nice Sophia Antipolis, UMR 6618 CNRS\\ 1361 route des Lucioles 06560 Valbonne - France} 
\author{J.P. Decruppe}
\affiliation{
Laboratoire de Physique des Milieux Denses, Universit\'e de Metz\\ 1 bvd Arago, 57078 Metz C\'edex 03}%

\date{\today}

\begin{abstract}

We report on the spatio-temporal dynamics of the interface in shear-banding flow of a wormlike micellar  system (cetyltrimethylammonium bromide and sodium nitrate in water) during a start-up experiment. Using the scattering properties of the induced structures, we demonstrate the existence of an instability of the interface between bands along the vorticity direction. Different regimes of spatio-temporal dynamics of the interface  are indentified along the stress plateau. We build a model based on the flow symetry which qualitatively describes the observed patterns. 
\end{abstract}

\maketitle
Many complex fluids of various microstructures strongly exhibit non linear properties under simple shear flows. When the characteristic times of the flow ($\simeq 1/\dot\gamma$) and of the system are similar, the system is liable to undergo instabilities and flow-induced phase transitions. Such coupling between flow and structure has been observed for example, in wormlike micellar solutions \cite{Ber3}, lyotropic lamellar phases  \cite{Sal2}, telechelic polymer networks \cite{Ber5}, soft glassy materials \cite{Cou1} or foams   \cite{Deb1}.\\ 
For a system showing a shear-thinning  behaviour, the mechanical signature, in the measured flow curve ($\sigma=f(\dot\gamma)$), of this type of phenomena is the stress plateau separating low and high viscosity branches. This stress plateau is at the origin of a shear-banding transition from a flow at first homogeneous towards a stationary state where two layers supporting different shear rates coexist. A change of the applied shear rate only affects the relative proportion of each layer, the width of the induced band increasing linearly with the macroscopic shear rate \cite{Cat1}.\\ Among the complex fluids with stress plateau, the systems of reversible giant micelles have been the object of the most intensive surveys (see \cite{Ber3} for a review). Very recently, the use of dynamic light scattering to measure velocity profiles brought confirmation of the very simple shear-banding scenario previously mentioned for the widely studied CPCl/NaSal solution  \cite{Sal1}. However, the authors mention the existence of temporal fluctuations in the highly sheared band. More complex pictures have also emerged with the development of time- and spatially-resolved techniques such as rapid nuclear magnetic resonance (NMR) velocimetry \cite{Lopez1} and ultrasonic velocimetry \cite{Becu}, with for example, the observation of periodic or quasirandom oscillations of the interface position driven by wall slip. Moreover, the flow birefringence, technique sensitive to molecular alignment, revealed complex kinetics of banding formation and strongly differing banding organizations from one sample to another \cite{Ler1}.\\ Some of these behaviors are reproduced, at least qualitatively, by several theoretical works, which predict fluctuating or chaotic flows, taking into account a coupling between flow and microstruture via the concentration or the micellar length \cite{Fiel2}. They also motivated a recent theoretical work, in which the author, using the nonlocal Johnson-Segalman model, studies the stability of planar shear banded flow with respect to small perturbations with wavevectors in the interfacial plane \cite{Fiel1}. An interfacial instability is predicted in the plateau domain for almost all shear rates, the interface profile being then modulated by a wave with wavevectors confined along the flow direction. 
In the present letter, we report on the formation kinetics of the banding structure and follow particularly the spatio-temporal dynamics of the interface between bands for a sample of cetyltrimethylammonium bromide (CTAB) and sodium nitrate (NaNO$_3$)\cite{Cap}. Direct visualizations of the gap of the Couette cell in the plane vorticity/velocity gradient and variations of the shear stress are recorded simultaneouly during a start-up experiment. Using the  scattering properties of the induced structures, we show that the interface between the two bands becomes unstable with wavevector in the vorticity direction. Different patterns of spatio-temporal dynamics of the interface are observed, depending on the applied shear rate. Using arguments based on the symmetry of the flow, we show that the interface position is described by a generic amplitude equation of Kuramoto-Sivashinsky type, which reproduces qualitatively the observed patterns \cite{Cros,Med1} .\\
\begin{figure}[t]
\includegraphics{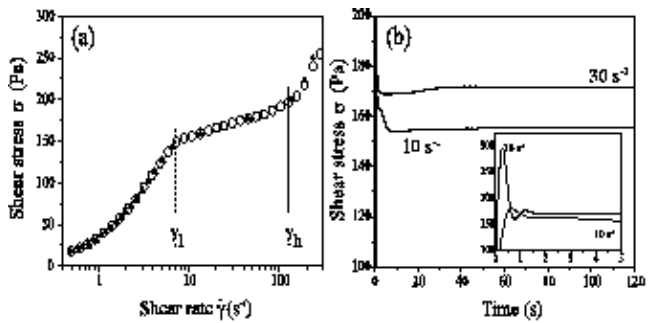}
\caption{\label{fig:epsart} (a) Steady state flow curves obtained under controlled strain rate ($\circ$, 250s per data point) and  controlled shear stress ($\blacktriangle$, 48s per data point). The dotted lines point out the bounded characteristic values of $\dot\gamma$ and $\sigma$ associated with the stress plateau.  (b) Variation of the shear stress versus time after the onset of flow for applied shear rates $\dot\gamma=10$s$^{-1}$ and $\dot\gamma=30$ s$^{-1}$. The inset shows the response on a shorter time scale.}
\end{figure}
The wormlike micellar system chosen in this work is made of CTAB at 11$\%$ wt. and NaNO$_3$ (0.405 M) in distilled water at T=30$^{\circ}$C. At this concentration, far from the transition concentration towards an orientated nematic phase at rest, the micelles form a highly entangled network. The linear viscoelastic behavior of the solution is Maxwellian with a plateau modulus $G_0=238\pm 7$Pa and a terminal relaxation time $\tau_R=0.15\pm 0.02$s. The rheological measurements are performed on a stress-controlled rheometer (Physica MCR500) working in strain-controlled mode. A home-build transparent Couette cell with a height of $40  \ \mathrm{mm}$ is used for the experiments. The inner rotating cylinder has a radius $R_1=13.33 \ \mathrm{mm}$ providing a gap thickness $e=1.13  \ \mathrm{mm}$. The Couette cell is designed in a way which allows the direct observation of the gap in the plane vorticity/velocity gradient. We form a thin laser sheet in that plane with a cylindrical lens and record the scattered intensity at 90$^{\circ}$ with a CCD camera.\\ 
The steady-state rheological behavior of our sample in strain- and stress-controlled modes of the rheometer is drawn in the semi-logarithmic plot in Fig 1.a. The flow curve is composed of two increasing branches separated by a stress plateau extending from $\dot\gamma_{l}=7\pm 0.5$s$^{-1}$ to $\dot\gamma_{h}=140\pm 10 $s$^{-1}$, characteristic of the emergence of heterogeneous flow of shear-banding type. This plateau presents a significant positive slope partly due to the non homogeneity of the Couette geometry. In fact, the curvature effects, analysed as in  \cite{Sal1} do not allow for a complete explaination of the slope. Concentration effects between shear bands could also make the plateau steeper \cite{Ovi1}.\\
We shall now focus on the kinetics of formation of the heterogeneous flow. At t=0, a shear rate is applied to the sample initially at rest and the evolution of the shear stress and direct observations of the gap of the Couette cell in the plane vorticity/velocity gradient are recorded simultaneously as function of time until steady-state is achieved. Fig. 1.b illustrates the variations with time of the shear stress for two shear rates associated with the plateau region ($\dot\gamma\geq\dot\gamma_{l}$). The stress $\sigma(t)$ follows the generic behavior observed on various semi-dilute micellar systems \cite{Ber1,Ler1} undergoing shear-banding transition induced by the flow: an overshoot at short times, then a slow sigmo\"idal relaxation  ($\dot\gamma=10s^{-1}$) or damped oscillations ($\dot\gamma=30s^{-1}$)) and finally a small undershoot that continues the transient response towards the stationnary state.\\
Let us correlate such an evolution with the direct visualizations of the gap (Fig. 2). The sample at rest does not present any particular scattering properties as observed on the first photo ($t=0^{-}$) where the gap appears dark. Just after the inception of flow, the solution becomes turbid over the entire gap and strongly scatters the incident light in all directions. This strong increase of the turbidity is associated with the elastic overshoot in the shear stress response (see photo 1). The entangled micellar network is strongly stretched leading to  concentration fluctuations. Then, the liquid becomes transparent again near the fixed wall of the  Couette cell (photo 2), this phenomenom correponding to the rapid relaxation of the stress overshoot. Let us note however, that the turbidity does not relax all over the gap since a scattering band persists against the rotating inner cylinder, the interface between bands being diffused (photo 4). The crucial point here, is that the induced structures are nucleated during the stress overshoot and in all the gap of the cell so as to the induced band is already formed just after the stress overshoot. The enhancement of the concentration fluctuations acts as the precursor of the transition. Both bands have differing optical properties: the induced band is very turbid, indicating that it contains structures of mesoscopic size which scatters the light in all directions. Such a difference allows us to perflectly distinguish the interface between bands.  Let us note that the turbidity is associated with birefringence banding.
\begin{figure*}
\includegraphics[scale=1]{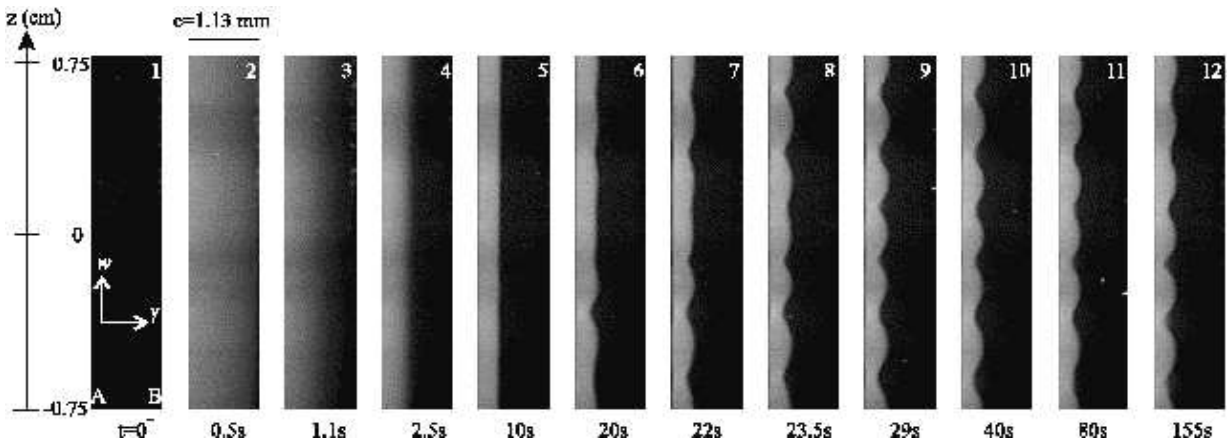}
\caption{\label{fig:wide} Views of the gap in the plane velocity gradient/vorticity taken out from the recording of the scattered intensity at different instants during the formation of the induced band. The sample is subjected to a shear rate of $30$s$^{-1}$. The letters A and B indicate  respectively the inner  and outer cylinders of the Couette device. Due to the compromise between the spatial resolution and the size of the field of observation, this latter is limited to 1.5cm in height and centered at halfway of the  cell. }
\end{figure*}
From $t=10$s  the interface becomes clearly sharp (photo 5). This lasts for about $t=20$s (photo 6) and then  we observe a completely unexpected behavior: the sharp interface between the two bands begins to destabilize itself along the vorticity axis. The instability grows up to $t=29$s (photo 9), namely when the stress undershoot occurs in the transient response. Beyond, the interface seems to adopt a stationnary periodic profile with a well-defined characteristic wavelength. An extension of the field of observation shows that this ripple of the interface spreads over all the height of the inner cylinder. At the edges of the cylinder, the ripple presents a minimum of amplitude.\\ This destabilization process of the interface occurs in all the plateau region of the flow curve. In a general way, the wavelength and the amplitude of the instability clearly increase with the applied shear rate, except when $\dot\gamma$ approaches $\dot\gamma_h$ where they diminish again. Let us note that above $\dot\gamma_h$, the gap is completely filled by the induced band.  Using a simple procedure of  image analysis, we detect the interface profile on each frame. We represent in gray levels, the amplitude of the interface as function of time and space coordinate along the vorticity axis as illustrated in Fig.3.a and b., where we summarized the complete  behavior of the interface on spatio-temporal diagrams for $\dot\gamma=10$s$^{-1}$ and $\dot\gamma=30$s$^{-1}$ respectively. 
In first approximation, we adjust the interface profile by a sinusoïdal function in order to estimate the wavelength and the amplitude of the instability. The observed patterns are very different for both the applied shear rates. At 10s$^{-1}$, the deformation of the interface occurs around $t=20$ s and the waves formed seem to propagate along the z-coordinate, alternatively towards the top and the bottom of the cylinder, interacting in a complex scenario. We find for the amplitude and the mean wavelength, respectively 0.007$\pm$0.003mm and 0.4$\pm$0.1mm.
The evolution at $\dot\gamma=30$s$^{-1}$ appears simplier since there is no propagation in this case. The interface keeps a spatially stable profile in course of time with an amplitude of 0.05$\pm$0.01mm and a wavelength of 2.4$\pm$0.02mm namely, more than two times the gap width. We checked by increasing the duration of the experiment up to 40 minutes that there is no coarsening between the "domains" (black and white) at very long times. A closer inspection of the diagram reveals the existence of time-periodic oscillations. We can distinguish two different frequencies: the first one, in the dark zones, is due to the mechanical imperfection intrinsic to the rheometer ; the second one, which is visible in the clear zones, is lower and results from the superposition of the default of coaxiality of the cell with a "beat" of the minimum of amplitude of waves. In other words, the amplitude of the interface instability oscillates with time. This could be due, among others, to tridimensional flow or to destabilization of the interface in the velocity direction \cite{Fiel1}.\\
Moreover, from the asymptotic behavior we are able to deduce the proportion of the induced "phase" $\phi_h$: the integration of the interface profile gives $\phi_h=0.29\pm 0.02$. The same calculus for each photo from $t=10$s gives the same relative proportion. According to the simplest shear-banding scenario described  by Cates's theory \cite{Cat1} and recently observed by Salmon et al \cite{Sal1}, the width of the induced band  increases linearly with the applied shear rate. In our case, the proportion of the induced band  computed from the lever rule seems to be underestimated : for $\dot\gamma=10$s$^{-1}$ and $\dot\gamma=30$s$^{-1}$, we find  respectively $\phi_h=0.02\pm 0.01$ and $\phi_h=0.18\pm 0.02$, values to be compared to $0.04\pm 0.01$ and $0.30\pm 0.02$. Processing of ultrasonic data on this sample is in progress in order to properly correlate the optical response and the local structure of the flow.\\
\begin{figure}[b]
\includegraphics[scale=1]{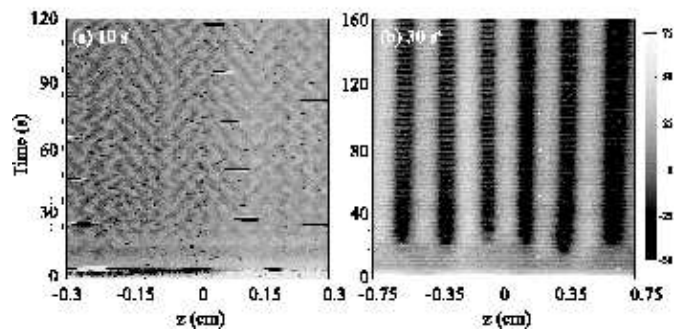}
\caption{\label{fig:epsart} Spatio-temporal evolution of the interface amplitude  for $\dot\gamma=10$s$^{-1}$ (a) and $\dot\gamma=30$s$^{-1}$ (b) on a time scale which provides both kinetics of destabilization of the interface and the asymptotic behavior. The z-coordinate gives the spatial position along the vorticity axis.  A linear gray scale (in $\mu m$) is used, the black and white regions being associated respectively with minima and maxima of the interface amplitude.}
\end{figure}
 We would like to propose a simple model that captures the asymptotic dynamics of the fluid flow. As stated previously, when the fluid is forced with a shear rate that is inside the plateau region, the system spontaneously coarse-graines into two zones separating two different shear rates.
 Here, as in lubrication approximation for newtonian fluids, gradients with respect of the transversal coordinate $y$ are higher than those with $z$ the coordinate tangential to the cylinder axis. This length scales difference yields that the dynamics with respect to $z$ coordinate are slow, and will consequently drive the interface position for asymptotic times. 
 Let us note $p$ the position of the interface separating these two zones. 
 Temporal evolution of the front must reflect the symmetry of the experiment $z\rightarrow -z$, so only terms including spatial derivatives like $\partial_{zz} p$, $\partial_{zzzz} p$ and $(\partial_{z} p)^2$ are allowed at lowest order. The system is not invariant with respect to translation in the $y$ direction, and the equation must present the broken symmetry : $p \rightarrow p+\mathrm{constant}$. The asymptotic dynamics of the interface are given by:
\begin{equation}
	\partial_t p=\mu (1-p)-\partial_{zz}p-\partial_{zzzz} p+\left(\partial_z p\right)^2.
	\label{KS}
\end{equation}
Equation \ref{KS} is written in non dimensional variables, and introduces the control parameter $\mu$ in which all the physical attributes of the experiment have been set.
\begin{figure}[t]
\includegraphics[width=\columnwidth]{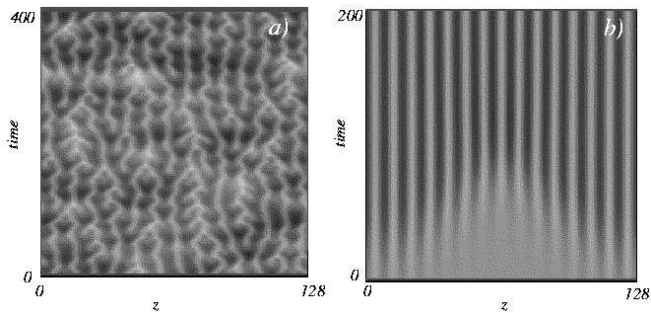}
\caption{\label{fig:ks} Spatio-temporal evolution of the interface position  $p$ obtained by numerical simulation of the model \ref{KS} with (a) $\mu=0.07$ (b) $\mu=0.2$.}
\end{figure}
The first  term of the r.h.s of the equation \ref{KS} just states that if the interface has a stationary location.
Since the fluid is set into negative viscosity inside the plateau, the second term is a diffusion term  with a negative sign. The next term, $\partial_{zzzz} p$  is a linear saturating term. The last term is the first non linear term that respects the symmetry at that order. This equation is known as the damped Kuramoto- Sivashinsky equation \cite{Chate}. When  $\mu>1/4$  the solution $p=1$ is a linearly stable solution. For $\mu<1/4$, the perturbed interface becomes oscillatory in space, with a well defined wave-length. Finally, for $\mu<0.078$, the dynamics are chaotic \cite{Chate} , see for example Fig. \ref{fig:ks} (a).
When $\mu=0$, equation \ref{KS} reduces to the usual Kuramoto-Sivashinky equation, and exhibits complex spatio-temporal behavior that is used to describe phase turbulence \cite{Kuramoto}. The existence of stationary periodic pattern observed in the experiment is related to the very low aspect ratio of the Couette cell. Let us finally note that, in the model, when the interface becomes unstable, its position is systematically shifted due to the non-linear term $(\partial_x p)^2$. This term has the same effect as the curvature of the Couette geometry. This means that the instability of the interface could also explain the "additional" slope of the plateau.	\\
In summary, our rheo-optical study of the interfacial dynamics in shear-banding flow reveals the existence of a destabilization process of the interface between bands with wavevector in the vorticity direction. This observation suggests that the flow field is three dimensional with the presence of recirculations. We show that there are different spatio-temporal dynamics of the interface along the stress plateau, with chaotic events at low shear rates, and spatially stable oscillatory behavior at higher strain rates. 
The physical origin of the destabilization process is still to be determined ; however, it could partly explain some fluctuating behaviors highlighted recently in shear banded flows \cite{Sal1,Lopez1,Becu} and it emphasizes the necessity for a complete spatio-temporal description of the flow and the microstructural organization in the different planes. Further, we buid a model equation \ref{KS} based upon symetry arguments which described qualitatively the observed complex patterns. This model equation can also be derived using perturbation theory. To this end, it is necessary to adopt a model for the rheology of the fluid \cite{Fiel1, Dhont}. It is a computation that we did  using the model \cite{Dhont} and we will report it elsewhere \cite{Lad} with the complete exploration of the stress plateau.\\ 
The authors are grateful to G. Gr\'egoire, O. Greffier, O. Cardoso, J.F. Berret, and J.L. Counord for the development of the  transparent Couette cell.


\begin{thebibliography}{10}
\bibitem{Ber3}
J.~F. Berret.
\newblock {\em Cond.-Matt.}, 0406681, 2005 and ref therein.

\bibitem{Sal2}
J.~B. Salmon, S.~Manneville, and A.~Colin.
\newblock {\em Phys.\ Rev.\ E}, 68:051503, 2003 ; J.~B. Salmon, S.~Manneville, and A.~Colin.
\newblock {\em Phys.\ Rev.\ E}, 68:051504, 2003.


\bibitem{Ber5}
J.~F. Berret and Y.~Serero.
\newblock {\em Phys.\ Rev.\ Lett.}, 87:4, 2001.

\bibitem{Cou1}
Coussot et~al.
\newblock {\em Phys.\ Rev.\ Lett.}, 88:218301, 2002.

\bibitem{Deb1}
Debr\'egeas.
\newblock {\em Phys.\ Rev.\ Lett.}, 87:4, 2001.

\bibitem{Cat1}
M.E. Cates, T.C.B. McLeish, and G.~Marrucci.
\newblock {\em Eur.\ Phys.\ Lett.}, 21:451, 1993.

\bibitem{Sal1}
J.~B. Salmon, A.~Colin, and S.~Manneville.
\newblock {\em Phys.\ Rev.\ Lett.}, 90:228303, 2003.

\bibitem{Lopez1}
W.M. Holmes, M.R. Lopez-Gonzales, and P.T. Callaghan.
\newblock {\em Eur.\ Phys.\ Lett.}, 64:274, 2003 ; M.R. Lopez-Gonzales, W.M. Holmes, P.T. Callaghan, and P.J. Photinos.
\newblock {\em Phys.\ Rev.\ Lett.}, 93:268302, 2004.

\bibitem{Becu}
L.~B\'ecu, S.~Manneville, and A.~Colin.
\newblock {\em Phys.\ Rev.\ Lett.}, 93:018301, 2005.

\bibitem{Ler1}
S.~Lerouge, J.~P. Decruppe, and J.~F. Berret.
\newblock {\em Langmuir}, 16:6464, 2000 ; S.~Lerouge, J.~P. Decruppe, and P.~D. Olmsted.
\newblock {\em Langmuir}, 20:11365, 2004.

\bibitem{Fiel2}
S.M. Fielding and P.D. Olmsted.
\newblock {\em Phys.\ Rev.\ E}, 68:036313, 2003 ; S.M. Fielding and P.D. Olmsted.
\newblock {\em Phys.\ Rev.\ Lett.}, 92:084502, 2004.

\bibitem{Fiel1}
S.~Fielding.
\newblock {\em Phys.\ Rev.\ Lett.}, 95:134501, 2005.

\bibitem{Cap}
E.~Cappelaere and R. Cressely.
\newblock {\em Collo\"id.\ Polym.\ Sci.}, 275:4, 1997.

\bibitem{Cros}
M.C. Cross and P.C. Hohenberg.
\newblock {\em Rev.\ Mod.\ Phys.}, 65:851, 1993.


\bibitem{Med1}
M. Argentina and M.G. Clerc and R. Rojas and E. Tirapegui.
\newblock {\em Phys.\ Rev.\ E}, 71:04210, 2005.


\bibitem{Ovi1}
O.~Radulescu and P.~D. Olmsted.
\newblock {\em J.\ Non-Newtonian.\ Fluid.\ Mech.}, 91:143, 2000.

\bibitem{Ber1}
J.~F. Berret, D.~C. Roux, and G.~Porte.
\newblock {\em J.\ Phys.\ II France}, 4:1261, 1994 ; C.~Grand, J.~Arrault, and M.~E. Cates.
\newblock {\em J.\ Phys.\ II France}, 7:1071, 1997.

\bibitem{Chate}
H.~Chate and P.~Manneville.
\newblock {Transition to turbulence via spatio-temporal intermittency}.
\newblock {\em Phys.\ Rev.\ Lett.}, 58(2):112, 1987.

\bibitem{Kuramoto}
Y.~Kuramoto.
\newblock {\em Chemical Oscillations, Waves and Turbulence}.
\newblock Springer, 1984.

\bibitem{Dhont}
J.K.G. Dhont.
\newblock {\em Phys.\ Rev.\ E}, 60:4534, 1999.

\bibitem{Lad}
S.~Lerouge, M.~Argentina, J.~P. Decruppe.
\newblock { in preparation.}

\end{thebibliography}
\end{document}